# Polar Coding and Linear Decoding

**Geraldo A. Barbosa**
KeyBITS Encryption Technologies LLC, Reston, VA 20190, USA
e-mail: geraldoabarbosa@keybits.tech

## Abstract

Polar encoding, described by Arikan in "IEEE Transactions on Information Theory, Vol. 55, No. 7, July 2009", was a milestone for telecommunications. A Polar code distributes information among high and low-capacity channels, showing the possibility of achieving perfect channel capacity. The high-capacity channels allow almost noiseless transmission of data. When these channels are not high noise, reliability is achieved in the signal transmission. It starts to compete against codes such a Low-Density Parity-Check (LDPC) codes.  Polar code can be also considered error correcting, based on the redundancy inherent in its structure. This feature makes polar encoding also applicable to digital quantum-resistant cryptography protocols.

This work explores <u>linear decoding</u> at a first or single trial in the case of small losses or small number of bit-flipping, and repeated transmission for medium level losses. This is distinct from Arikan's successive probabilistic decoding by application of probabilistic rules. Linear decoding is done directly from solving the <u>linear equations</u> connecting the codewords $x$ and the received signals $y$ after transmission via noisy channels. Numerical examples will be shown. Along with this work, programming in Mathematica language was used. Codes are available for copy-and-paste for Mathematica users to immediately try the described formalism.

## Introduction

While the coding stage of Arikan's successive decoding has a simple deterministic mathematical structure, decoding is probabilistic and much more involved. Several advances were needed for Polar decoding to be reliably employed in telecommunication levels. Decoding complexity for large kernels often leads to inefficient implementations. Nevertheless, adoption of Polar codes in 5G by Huawei in 2016 achieves speeds of 27Gbit/s, nearing Shannon's limit and revealing the potential of the technique to achieve channel capacity with polynomial complexity.

Linear decoding is explored in this work and numerical simulations were implemented in a laptop computer to allow simple conclusions and, eventually, to stimulate experiments with faster hardware techniques such as use of FPGA or ASIC. This works starts by defining notations used in this work and concisely presents Arikan's polar coding. The ideas of Binary Erasure Channel and Gaussian bit-flipping due to transmission channel noise are discussed, with numerical examples. The linear decoding codes are shown in the Mathematica programming language, applied, and results are discussed.

## Signals in the telecommunication channels

A sketch of signals in telecommunication channels is introduced to make connections with notations used in this work and Polar coding operations [1].

Figure 1 introduces some of the notations used in this paper. It sketches a communication line, where a message $u_i$, (sequence of binary symbols $i = 1,2,3, … N$) in a transmission station TX is encoded for transmission to the network. The noiseless coded signals are represented by $x_i$. After transmission by the noisy channel, received signals $y_i$ are $x_i$ degraded by noise. At the receiving station RX, $y_i$ must be decoded and the message $u_i$ recovered:

message $\boldsymbol{u}$ (=set of $u_i$ as a vector) → noiseless coded words $\boldsymbol{x}$ (=set of $x_i$)
→ received noisy signals $\boldsymbol{y}$ (=set of $y_i$).

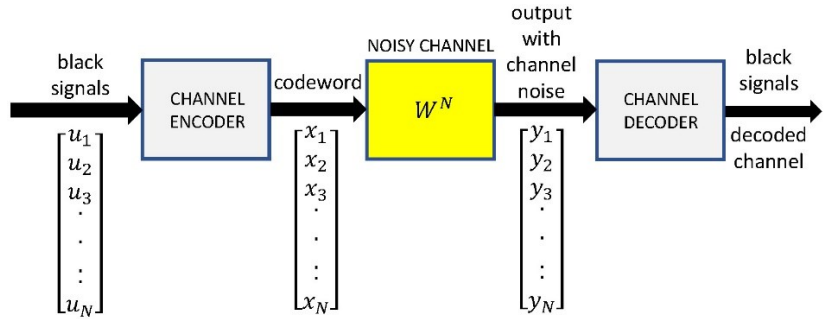

*Figure 1 – Communication line (starting in a transmission station TX), where the message $u_i$ (black arrow), that may have aggregated error correcting codes, inputs the channel encoder (for ex., given by polar coding) and outputs the codewords $x_i$. The codewords pass by the noisy transmission network ("noisy channel") and arrive at the receiving station as the sequence of signals $y_i$. The encoding operations are deterministic and, as such, given $x_i$, can be reverted to $u_i$. The channel transmission and signal $y_i$ are noisy when arriving at the receiver. In principle, the channel decoder should use the properties of the coding protocol and, together with the error correcting codes, allow recovery of the message $u_i$. For quantum-resistant protocols, $u_i$ may represent signals generated by the these protocols.*

Polar coding steps, in Mathematica language, are shown in Appendix "Polar Coding". In Figure 1, $\boldsymbol{u}$ is the open message to be encoded by the Channel Encoder generating the codeword $\boldsymbol{x}$.

Polar encoding [2] uses a particular form of channel splitting, that can be written in algebraic form. After transmission via the noisy channel, explicit dependences are determined for the loss parameter $\epsilon$ in the case of a BEC (Binary Erasure Channel) model, or an AWGN (Additive White Gaussian Noise) channel. Explicit dependences are useful, due to time-varying losses in some networks, frequency dependencies, allowing quick reconfiguration of channels. Loss probes or pilot signals in real telecommunication channels may extract the losses (such as $\epsilon$ in BEC or $\sigma$ values in AWG channels) at different moments, feedbacking the *coding* process to adjust to the varying $\epsilon$ or $\sigma$ values.

## Polarization of Mutual Information

Observing polarization of the channels using Arikan's vector decomposition is useful to understand the transformation of message $\boldsymbol{u}$ (=set of $u_i$ as a vector), onto the noiseless coded words $\boldsymbol{x}$ (=set of $x_i$), shown in Appendix "Polar Coding".

Using a successive Martingale decomposition, Arikan shows that successively splitting a lossless information $u$ into two pieces of lossy information constrained to have their sum recovered to $u$, will lead to several channels with their Mutual Information function (channel capacity, for symmetric channels) covering the range $0 \leq I \leq 1$.

## Martingale decomposition

Mutual Information equations are synthesized by Arikan's Eq. (6) in Ref. [2]. Arikan's "Polarization" of the Mutual Information begins with the splitting of the first channel into two channels ("daughter" channels), where loss of a symbol (Binary Erasure Channel or BEC) has absorption probability $\epsilon$ and transmission probability $P = 1 - \epsilon$. The Mutual Information $I$ value 1 implies the perfect transmission of the bit (transmitter and receiver acquire the complete information), and 0 implies a complete loss.

It is emphasized that the Martingale decomposition with Mutual Information is achieved through a single "bit" $u$, while the full encoding scheme applies simultaneously to all bits of a specific communication with $N$ bits: $\boldsymbol{u} = u_1, u_2, \ldots u_N$. Distinct bits are intertwined in the expansion procedure. In other words, the encoding protocol (an expansion method) in the Martingale expansion for a single bit is not strictly equivalent to the multi-bit expansion.

Arikan's Polarization starts with the first channel with Mutual Information $I_0 = 1 - \epsilon$. The operation proceeds by splitting the first Mutual Information channel into the "daughter" channels with this first generation giving $I[2] = \{I_1, I_2\}$, where the Mutual Informations $I_1 = 1 - \epsilon^2$ and $I_2 = (1 - \epsilon)^2$. The Mutual Information for these channels obeys $I_1 > I_2$. Their sum is $I_1 + I_2 = 2 I_0$. In other words,

$$I_0 = \frac{1}{2}(I_1 + I_2), \tag{1}$$

where the equality applies $iff$ $I_0 = 0$ or 1. The specific dependence form on $\epsilon$ given to $I_1$ and $I_2$ is not crucial, as long as the splitting gives two Mutual Information functions: one of smaller value and the other with a higher value, and only depending on the last value used, defining a Martingale process, as illustrated in Figure 2, Figure 3, and Figure 4.

This note will use Mathematica's commands to facilitate immediate use for those familiar with this software. The decomposition procedure for the Mutual Information $I$ is defined by all subsequent channel splitting (doubling), generating the recurrence rule for an arbitrary $N$:

$$I[k\_] = \text{Flatten}[\text{Table}[F[k,s], \{s, 1, k\}]], \text{ where} \tag{2}$$

$$F[k,s] = \left\{2I\left[\frac{k}{2}\right][[s]] - \left(I\left[\frac{k}{2}\right][[s]]\right)^2, \left(I\left[\frac{k}{2}\right][[s]]\right)^2\right\} \tag{2a}$$

This recurrence rule gives a duplication of terms in each "generation", for example

$$I[2] = \{\{1 - \epsilon^2\}, \{(1-\epsilon)^2\}\}, \tag{3}$$

$$I[4] = \{\{2(1-\epsilon^2) - (1-\epsilon^2)^2, (1-\epsilon^2)^2\}, \{2(1-\epsilon)^2 - (1-\epsilon)^4, (1-\epsilon)^4\}\},$$

$$I[8] = \{\{2(2(1-\epsilon^2) - (1-\epsilon^2)^2) \qquad (4)$$
$$- (2(1-\epsilon^2) - (1-\epsilon^2)^2)^2, (2(1-\epsilon^2) - (1-\epsilon^2)^2)^2\}, \{2(1-\epsilon^2)^2 - (1-\epsilon^2)^4, (1-\epsilon^2)^4\}, \{2(2(1-\epsilon)^2 - (1-\epsilon)^4) - (2(1-\epsilon)^2 - (1-\epsilon)^4)^2, (2(1-\epsilon)^2 - (1-\epsilon)^4)^2\}, \{2(1-\epsilon)^4 - (1-\epsilon)^8, (1-\epsilon)^8\}\},$$

and so on.

The number of channels $N$ is chosen by a balance between benefit, cost, and the transmitted data length. A large $N$ produces a better splitting of the polarizing channels with an increased computational cost. For example, 5G technology has a standard of $N = 1024$ channels.

Each splitting represents distinct channels with losses. As the number of splittings increases, the polarization effect, described by Arikan, becomes evident. See these Martingale processes in Figure 2 to Figure 4, for the value $\epsilon = 0.5$.

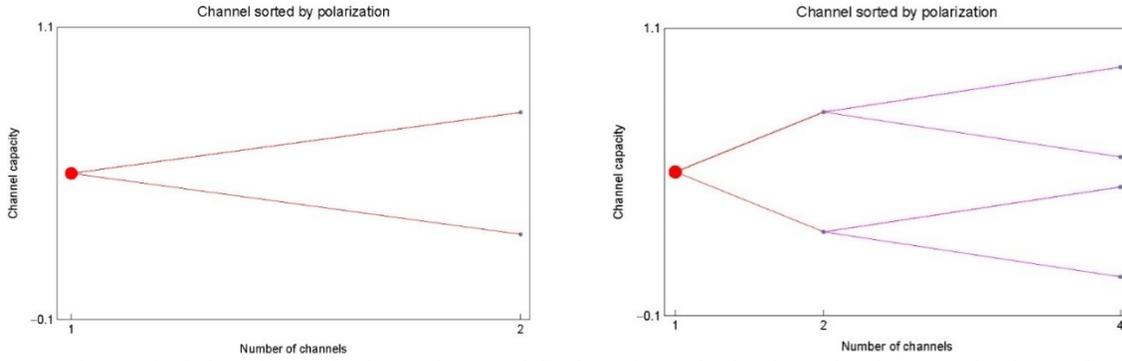

*Figure 2 – At the left is shown the first splitting of the first channel. On the right, the second splitting from each former channel.*

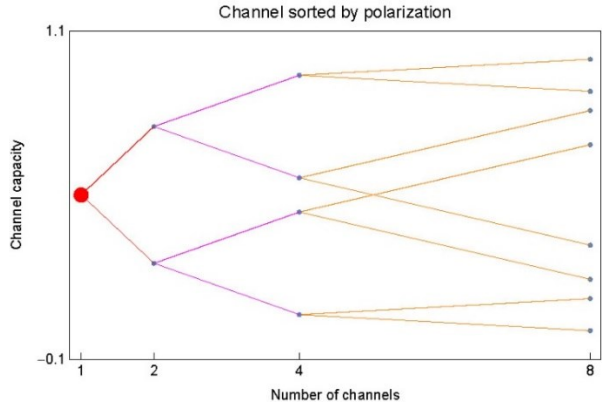

*Figure 3 – Third splitting giving $8 = 2^3$ channels*

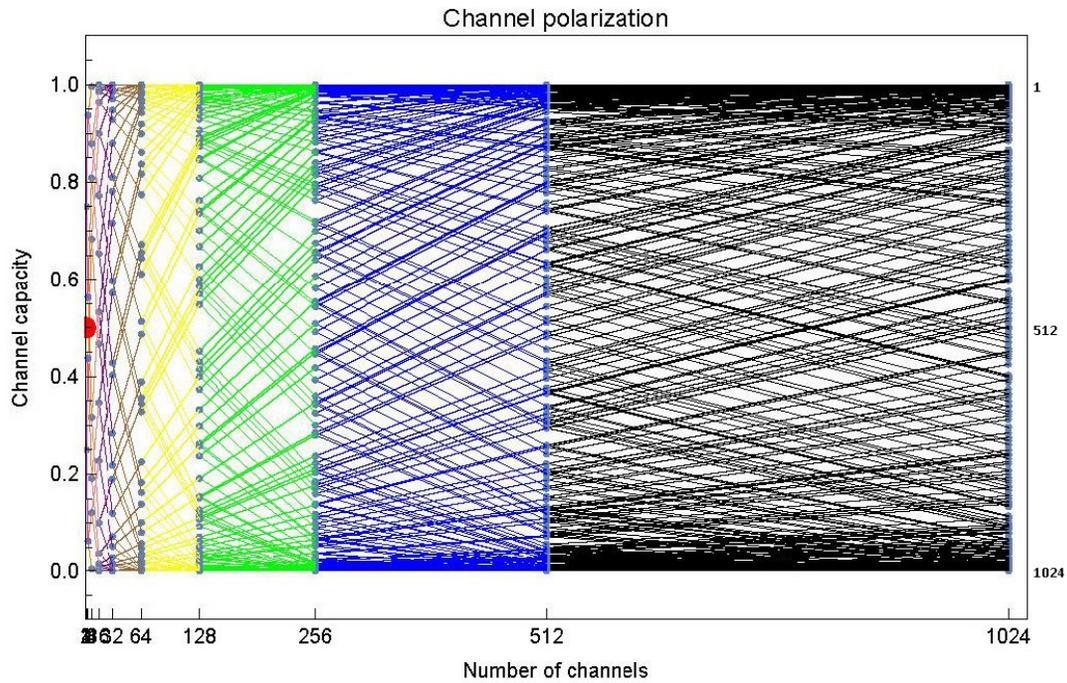

*Figure 4 – Mutual Information splitting up to $2^{10} = 1024$ channels, a Martingale process. The "condensation" of the channels towards noiseless and noisy channels as the splitting occur is evident –see the left vertical axis. The vertical axis at right shows the channel numbers (coding channels 1 to 1024, from top to bottom), where the horizontal axis shows the evolution of channels as the splitting occurs.*

## Channel density and splitting

The polarization effect shown in Figure 4 can be represented as a function of the sorted coding channels numbers, $N_s$, as indicated at the right side of Figure 5. It can be also represented as a function of the number of channels, $N$, as polarization is developed.

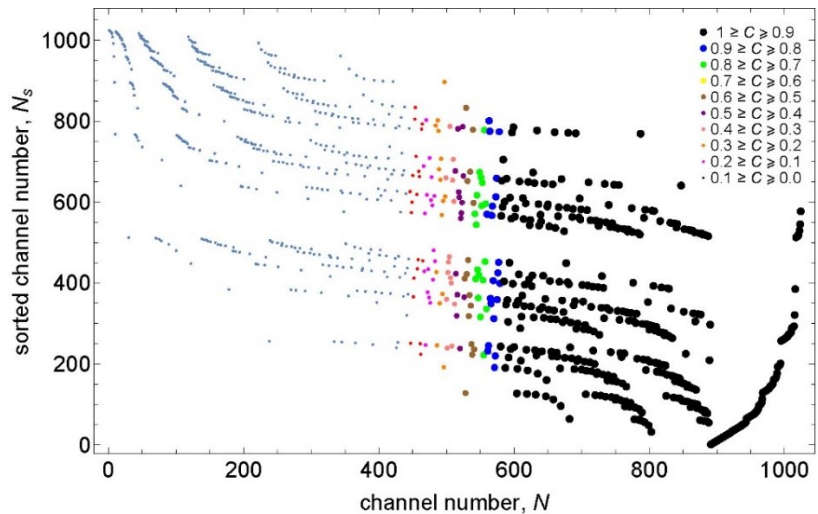

*Figure 5 – Representation of the polarization channels as a function of N and $N_s$. The dot sizes (with different colors) represent distinct Mutual Information levels, or channel capacity C. In this plot, only a coarse grain aspect is presented, with points grouped in C intervals of 0.1, as indicated on the top right inset.*

The same splitting, showing polar condensation, with increased resolution for the $C$ values, is seen in the 3D plot of Figure 6 (the adopted color representation is different in Figure 5 and Figure 6).

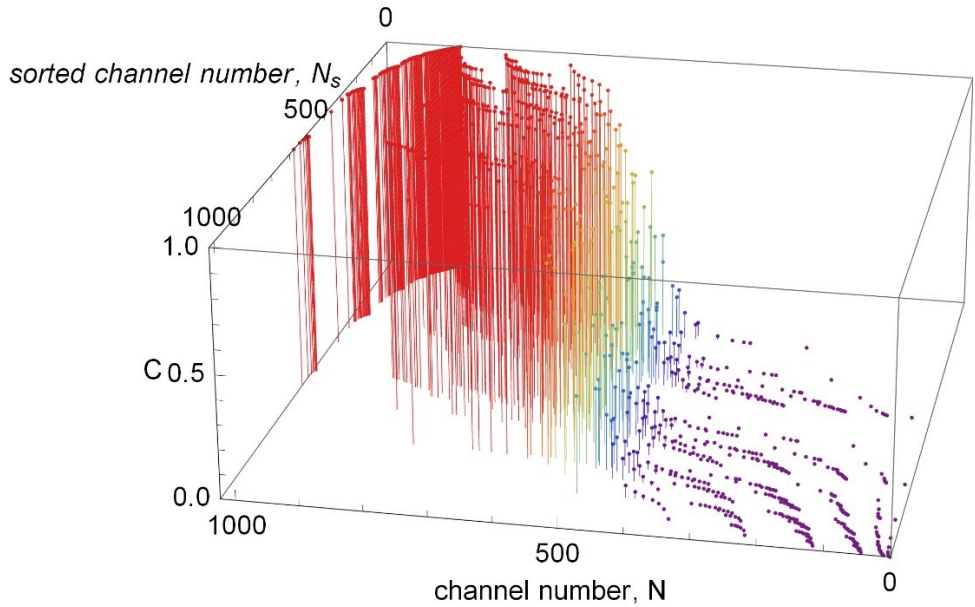

*Figure 6 – Capacity C, as the splitting occur, in 10 steps, from $2^1$ to $2^{10} = 1024$: Frontal plane "channel number, N", showing the polarization increase as the splitting occurs. N is the channel coding (coded binary output). The lateral plane with "sorted channel number, $N_s$", is where the channels with high value of the capacity C were sorted (High C values starting at $N_s = 1$). $N_s$ is the sorted channel coding (coded binary output).*

Lateral projections of Figure 6 are shown in Figure 7 and Figure 8.

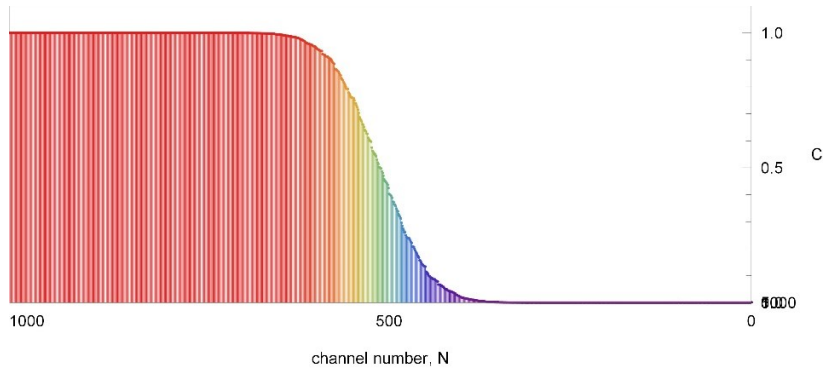

*Figure 7 – Channel polarization as a function of the number of channels N.*

A visualization of the sorted polarized channels numbered $N_s$ is shown in Figure 8.

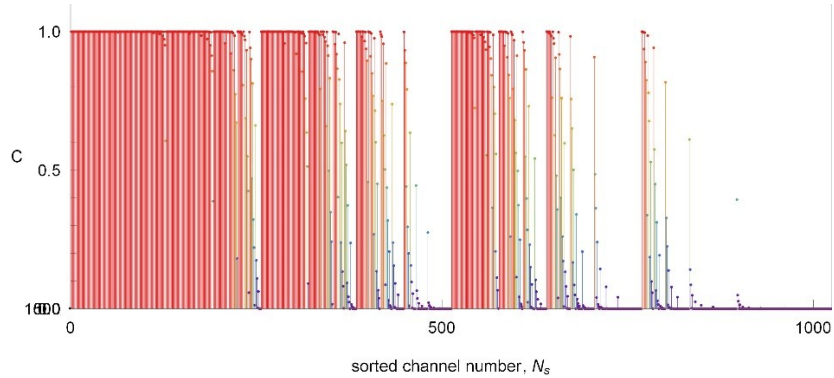

*Figure 8 - Channel polarization as a function of the number of channels $N_s$.*

## Histogram

A histogram of up to 1024 channels is shown in Figure 9. The agglomeration of channels around the maximum Mutual Information $I$ (or channel capacity $C$) values of $I = 1$ and $I = 0$ is evident.

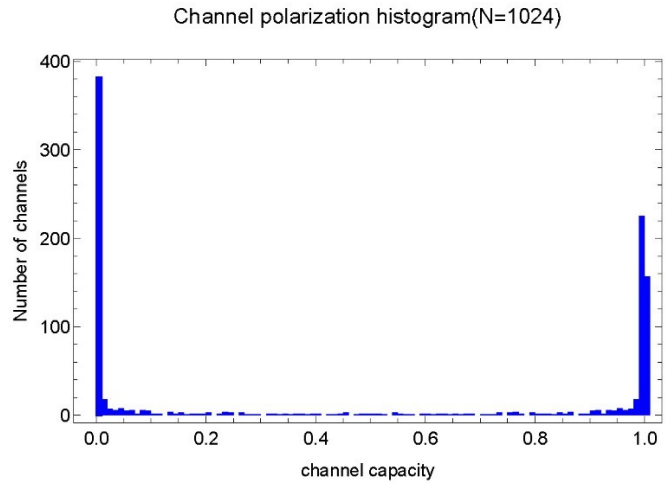

*Figure 9 – Histogram of the channel capacity for N=1024*

This polarization effect is represented in another way in Figure 10 (blue points). The horizontal axis represents the number of the ordered (or sorted) channel according to channel capacity. Capacity is represented in the vertical axis.

In fact, this distribution fits well as a Logistic Sigmoid function (see, for example, https://en.wikipedia.org/wiki/Sigmoid_function), as given by

$$f_{LS}(n) = \frac{1}{1 + e^{-\frac{n-\mu}{\beta}}}$$  (5)

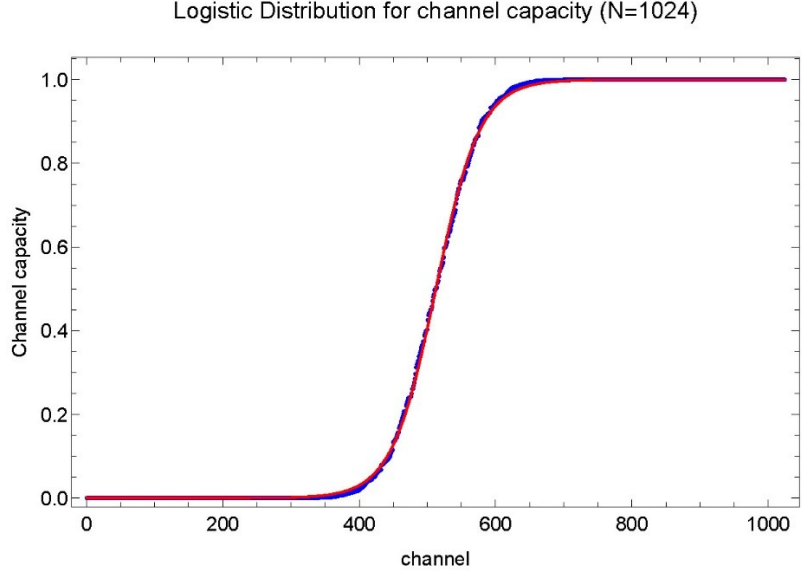

*Figure 10 – Distribution of Channel capacity.*

The data, given by blue points (See Figure 10), fit to $f_{LS}(x)$, using parameters $\mu = 512 \pm 0.07$ and $\beta = 3.14 \pm 0.06$. The theoretical line is the solid thin red line.

A channel density distribution can be derived from Eq. (7), assuming that density is proportional to the inverse of the derivative of the distribution $f_{LS}(x)$:

$$D_f = \frac{const}{\frac{df_{LS}(n)}{dn}}. \tag{6}$$

It can be normalized by $const \times \int_0^N D_f(n)dn = N$, giving the normalized form $D_{norm}(n)$:

$$D_{norm}(n) = \frac{N\left(1+\cosh\left[\frac{n-\mu}{\beta}\right]\right)}{N+\beta\left(\sinh\left[\frac{N-\mu}{\beta}\right]+\sinh\left[\frac{\mu}{\beta}\right]\right)}. \tag{7}$$

Figure 11 shows $D_{norm}(n)$ for $N = 1024$ channels. Low capacity and high capacity are symmetrically distributed with respect to $n = \mu = 512$.

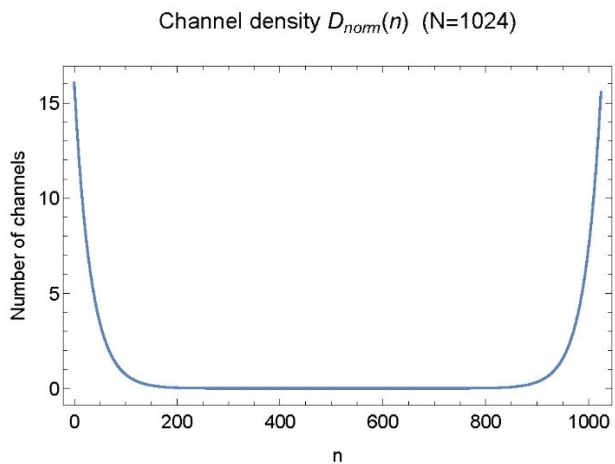

*Figure 11 – Channel density $D_{norm}(n)$*

# Decoding and probabilistic losses

The use of a fixed erasure parameter $\epsilon$ is an idealization. Other models of noisy channel exist such as AWGN model. These two models are briefly discussed ahead. In a real situation the most adequate model can be chosen to simulate the noisy channel.

## BEC channel

The Binary Erasure Channel (BEC) describes a transmission channel affected by noise in such a way that bits, 0 or 1, can be transmitted noiseless or lost due to absorption.

It is understood that $\epsilon$, that represents loss, is a probabilistic parameter. That is, bits are lost with average probability $\epsilon$. Small refinements can be made on the BEC model: Although specific networks may have good stability of losses, in general, one may refine the losses using an average loss $\bar{\epsilon}$ with some statistical deviation $\sigma_\epsilon$.

Time-dependent variations may also exist, but this will be ignored for now. In the same way described above, frequency $\nu$ modulations can be used to send signals through distinct channels. Loss parameters depend on frequency as well: $\epsilon(\nu)$. Very basically, 5G operational frequencies are set along a minimum of the atmosphere absorption, such as 28GHz, 38GHz, 58GHz and so on (see Figure 12).

Inclusion of all the dependences increases the decoding difficulties, because decoding uses probability parameters. These may be changing as communication pathways may be changing. One of the less-controlled losses is eventual shadowing due to physical obstructions along the propagation paths. However, atmospheric absorption is the most important and permanent condition.

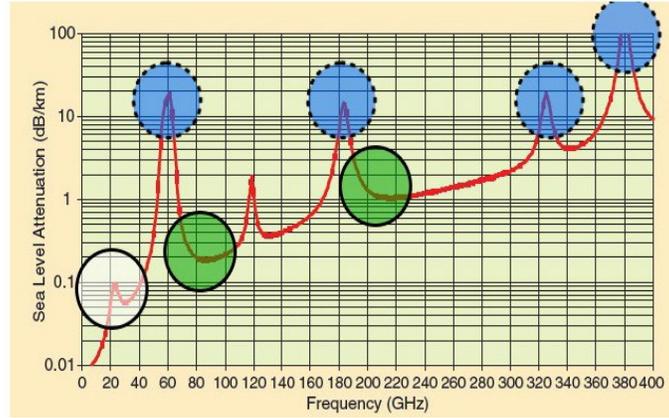

*Figure 12 – Air attenuation (extracted from MacCartney, G.R., Zhang, J., Nie, S., Rappaport, T.S., IEEE Global Communications Conference, Exhibition & Industry Forum (GLOBECOM), Dec. 9~13, 2013).*
*Air attenuation at different frequency bands is displayed. The white circle shows attenuation peaks at 28 and 38 GHz. The green circles show the attenuation around current communication systems, comparably larger than the white circle. The blue circles indicate frequencies with high attenuation, but which may be viable for low-range indoor communication or out of atmosphere.*

In telecommunication systems using polar codes, the erasure probability $\epsilon$ (for a BEC model) typically ranges from 0.001 to 0.2, depending on the system:

- High-quality channels (e.g., optical, high-SNR 5G): $\epsilon$ (0.001–0.01)
- Moderate channels (e.g., 5G NR control channels, VLC): $\epsilon$ (0.01–0.1)
- Low-SNR or noisy channels (e.g., wireless at 0–2.5 dB SNR): $\epsilon$ (0.1–0.2)

The idealized BEC channel, with a constant erasure loss, is shown in Figure 13.

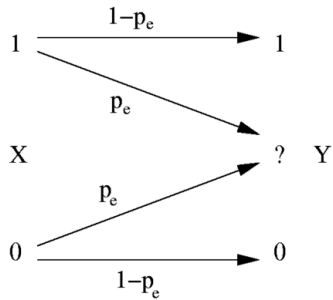

*Figure 13 – BEC with a constant erasure loss $p_e$ ($= \epsilon$). X is the emitter and Y is the receiver.*

The BEC can be characterized by

$$P[Y = 0|X = 0] = 1 - p_e \qquad (8)$$
$$P[Y = 0|X = 1] = 0$$
$$P[Y = 1|X = 0] = 0$$
$$P[Y = 0|X = 1] = 1 - p_e$$
$$P[Y = e|X = 0] = p_e$$
$$P[Y = e|X = 1] = p_e.$$

$p_e$ usually replaces $\epsilon$. The index $e$ represents an error.

It can be shown that

$$I[X;Y] = 1 - P[Y = e|X = 0] = p_e. \tag{9}$$

### Number of erased channels
The number of erased channels within the original $N$ channels can be estimated as

$$\text{Number of erased channels} = \text{Round}[N \times \epsilon]. \tag{10}$$

The choice of which ones are erased can be determined by the uniform probabilistic choice of $N \times \epsilon$ integers within $N$:

$$\text{ErasedChannels} = \text{RandomInteger}[N, \text{Round}[N \times \epsilon]] \tag{11}$$

## Decoding evolution and patents
Many Polar decoding activities focused on ways to achieve better efficiency. Both software and hardware solutions have been proposed.

Arikan's seminal work [2] was essential as the basis for decoding. Reviews [1] and detailed references [3] are available on the web. Speed is precious for telecommunications and hardware implementations. See, for example, references [4, 5, 6, 7, 8, 9, 10, 11, 12, 13], [14, 15, 16].

## BEC model. Coding and Decoding steps in the linear approach to Decode
The following describes the processes used for BEC[1] coding and decoding.

### Coding Steps
1) Starting from the sender, a message with a size $N/2$ is created, mssg= Length$[\frac{u}{2}]$, which is a vector with binary components $u_i$ $(i = 1,2, \dots N)$.

2) Create a binary sequence to be used as "frozen" bits, and not coincident with the data bits. In this example the number of frozen bits has the same size as the message $(= N/2)$.

3) Join the frozen bits with the message bits:

$$\text{Join}[\text{frozen}, \text{mssg}] \equiv uv \text{ (size } N).$$

$uv$ is the packet to be coded giving $x$ (= "codeword", another vector). $x$ will be the coded message formulated from the polar coding expansion of $u_i$.

4) Define a random noise model, such as BEC, Gaussian, or other, to simulate the noisy transmission channel. In this example, BEC is used. Introduce noise $n$ onto $x$, to create

---
[1] see https://en.wikipedia.org/wiki/Binary_erasure_channel

### Linear decoding of the BEC channel

5) The coded **x** has a specific form, <u>known to both sender and receiver</u>. Create an algebraic sequence $y[k]$ with the same structure of $x[k]$ ($x[k]$ is made from combinations of $u[j]$, with ($j = 1,\ldots k$) but with another name for the bits, such as $y[ur[j]]$.

$ur[j]$ will contain combinations of message bits, frozen bits. The received number of $ur[j]$ will be reduced from $N$ due to the erased bits in the transmission channel.

6) Use "Reduce" to solve for bits $ur$:

$$\text{Reduce}[\ y[ur] == yreceived,\quad ur]. \tag{11}$$

$yreceived$ contains the frozen bits $f$ (at known positions) as well as empty values due to bit losses due to $\epsilon$.

The correct decoding results in $ur \rightarrow u$, which recovers the message bits from the noisy signals.

Some examples are presented below and after these examples detailed explanations are given.

### Examples of time-decoding for $N = \{8,16,32,64,128,256,512,1024,2048,4096\}$

Figure 14 exemplifies successful decoding times in a single-trial for a low loss channel with $\epsilon = 0.01$. This extracts a dependence of decoding time as a function of $N$.

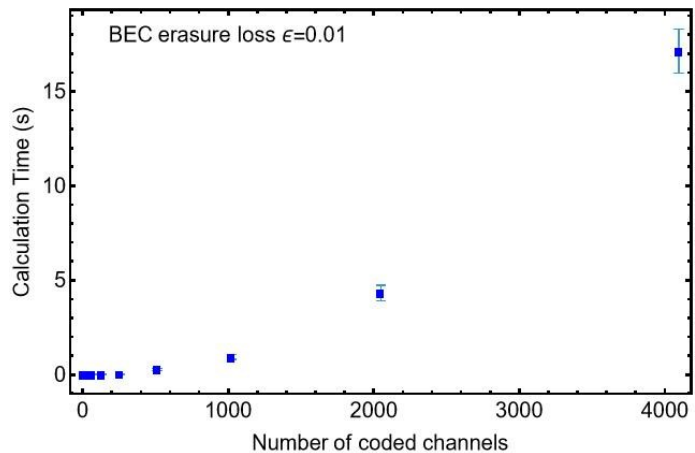

*Figure 14 – Decoding times as a function of N for BEC=0.01. Each point is a result of multiple coding-decoding. The error bars show the standard deviations.*

Figure 15 shows a fit of the dependence of the decoding time versus N. It fits with a $N^2$ dependence:

$$\text{Coding/Decoding calculation time}[N] \sim 10^{-6} N^2 \text{ (in seconds)} \tag{12}$$

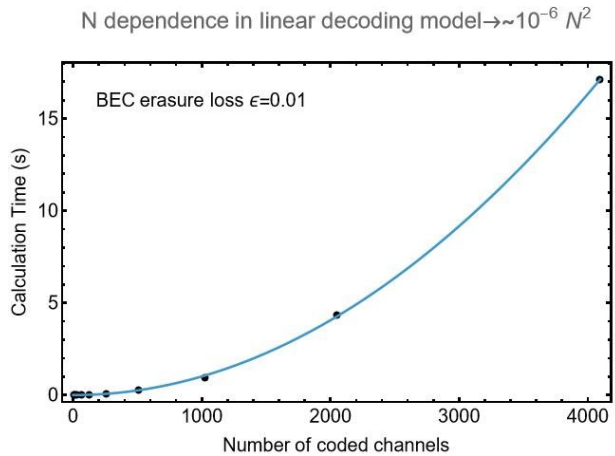

*Figure 15 – The dependence of the decoding time versus N. The black points are data. The continuous curve is the fit model $10^{-6}N^2$*

## Arikan's Encoding protocol

### Encoding diagrams

The communication diagrams represent $N$ inputs $u_i$ ($i = 1,2,3,...N$) and $N$ outputs $r_i$ ($i = 1,2,3,...N$), of a communication system. The inputs may be understood as representing information bits $u_i$ injected at a transmission station TX, codified or "encoded" in a deterministic way to symbols $x_i$, and then sent to a receiving station RX, arriving as signals $r_i$. See Figure 16.

The in-transit operation between TX and RX is assumed lossy, represented by the box $W$ in the communication diagram. The encoding part is *lossless,* representing deterministic operations, It includes all steps from the input $u_i$ to the $W$ boxes, which includes $XOR$ gates and connection combinations. See Figure 16 through Figure 19.

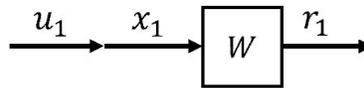

*Figure 16 – Single channel with input $u_1$ and output $r_1$. In this simple case, no operations are represented between the input $u_1$ and the encoded signal $x_1$ to be transmitted through the lossy channel $W$.*

Figure 17 is the very basic step in this protocol:

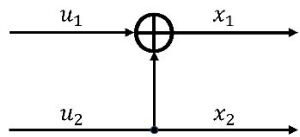

*Figure 17 - Two inputs, $u_1$ and $u_2$, one gate XOR, and the resulting encoded signals $x_1$ and $x_2$.*

It results from the application of matrix operation

$$F_0 = \begin{pmatrix} 1 & 0 \\ 1 & 1 \end{pmatrix} \quad (13)$$

to the input vector $\{u_1, u_2\}$:

$$\{u_1, u_2\}. F_0 = \{u_1 + u_2, u_2\}, \tag{14}$$

where the + operation part is understood as $\text{Mod}[u_1 + u_2, 2]$. This is an XOR operation.

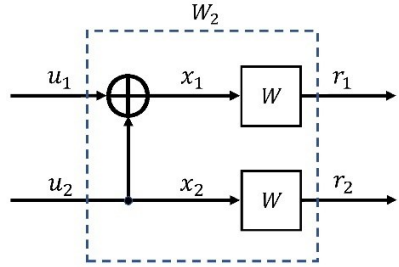

Figure 18 - Two inputs, $u_1$ and $u_2$, one gate XOR, and the resulting encoded signals $x_1$ and $x_2$, two lossy channels W and two outputs $r_1$ and $r_2$.

Figure 19 shows a four-bit input and permutations among groups of bits.

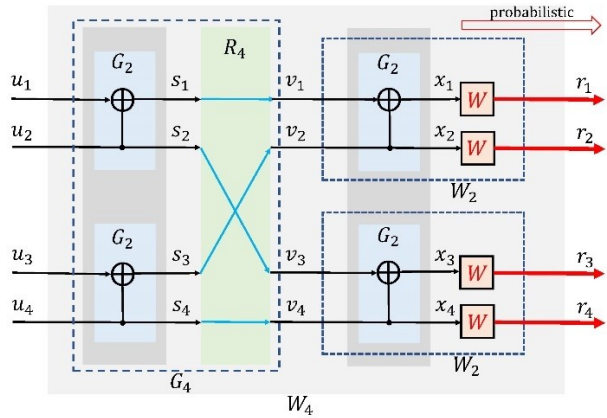

Figure 19 - Four inputs, $u_1$ to $u_4$, four gates XOR, symbolized by $G_2$, a routing gate $R_4$, that makes permutations among channels, and among the resulting encoded signals $x_1$ to $x_4$, four lossy channels W and four outputs $r_1$ to $r_4$. Processes from the $u_i$ inputs to the encoded signals $x_i$ are determinist operations. The encoding step occurs in TX. Blocks $G_2, G_4, W_2$ and $W_4$ represent operational modules that will be used to simulate the signal space with N channels.

The $R_N$ operation is a reverse-permutation of indexes. $B_N$ (not in these figures), is a bit reversal. Operations $R_N$ and $B_N$ will be described later.

## Lossy processes

The objective of the polar coding technique is creating channels that maximize efficient transmission of information. This is achieved by splitting the information among multiple channels, where some channels are more lossy than others.

*Losses* in the BEC model are represented by a single parameter $\epsilon$: symbols lost during transmission: in a percentage $\epsilon$ of the total of symbols sent, a bit 0 or 1 may be lost by signal absorption or other reason. The transmitted fraction is 1- $\epsilon$.

Losses can be due to many factors, controlled or not. For example, the range of frequencies used in 5G is broad, ranging from sub-GHz to hundreds of GHz. Sources of losses can be traced to many factors, including different modulation techniques, transmissivity conditions, specific modulators, demodulators and detection equipment. In addition to signal communications, geographic network placement, variability of the transmission and reception places and antenna placements are a few of many other factors that can contribute to losses.

In principle, low loss channels are chosen to send messages, and the lossy channels are just filled with redundant information that can be neglected at the end. This splitting is described after the discussion of encoding and decoding. *Encoding* and *decoding* constitute the polar *code* protocols.

## Fundamental encoding operations

The $R[N]$ gate makes a <u>permutation</u> of a bit sequence to be transmitted into two-bit groups where the first group in the sequence has odd indexes and the remaining one carries even indexes. For example, $R[4]$ maps $\{s_1, s_2, s_3, s_4\}$ into $\{s_1, s_3, s_2, s_4\}$.

$B[N]$ is a successive application of $R[N]$ in steps of 2: $B[N] \equiv R[N]\left(I_2 \otimes R\left[\frac{N}{2}\right]\right)\left(I_2 \otimes R\left[\frac{N}{2}\right]\right) \dots (I_2 \otimes R[2])$, where $I_j$ are identity operators in dimension $j$. $B[N]$ is a bit reversal operator. This gives (Arikan's Eq. (71)):

$$B[N] \equiv R[N]\left(I_2 \otimes B\left[\frac{N}{2}\right]\right). \tag{15}$$

Application of $F_0$ in a "$n$"-nested operation is defined by

$$F^{\otimes n} := \text{Nest}[\text{KroneckerProduct}[\#, \#]\&, F_0, n-1]. \tag{16}$$

Identification with Arikan's $G[N]$ is made by $G[N] = K[N] = F^{\otimes n}$, if no permutation is applied.

$B[N]$ applies permutations and leads to the $G_p[N]$, the permuted $G[N]$ form:

$$G_p[N] = B[N]F^{\otimes n}. \tag{17}$$

This is Arikan's Eq. (70). Notation $G[N]$ and $G_p[N]$ are not standard and should be checked against papers in the literature.

The sequence of operations involving $R[N], B[N]$ are represented by the final operator $G[N]$, that upon the input $u[N] = \{u_1, u_2, \dots u_N\}$ produces the encoded signal $x[N] = \{x_1, x_2, \dots x_N\}$:

$$x[N] \equiv \text{Mod}[u[N], G[N], 2]. \tag{18}$$

The fundamental operation in $G[N]$ is an $XOR$ gate, shown before, produced with

$$F_0 \equiv \begin{pmatrix} 1 & 0 \\ 1 & 1 \end{pmatrix}$$

that, when applied to two inputs $\{u_1, u_2\}$ gives

$$\{u_1, u_2\}.F_0 = \{u_1 + u_2, u_2\}.$$

This should also be understood as followed by a modulus 2 operation, Mod[,2], giving

$$\text{Mod}[\{u_1 + u_2, u_2\}, 2].$$

For binary inputs, the XOR operations will always be assumed.
To obtain the final operator $G[N], N > 2$, one needs the expansion of $F_0$ to power $\text{Log}[2, N]$:

$$F[2] = \text{KroneckerProduct}[F_0, F_0], \quad F[3] = \text{KroneckerProduct}[F_0, F_0, F_0], \ldots$$

$$F[10] = \text{KroneckerProduct}[F_0, F_0, F_0, F_0, F_0, F_0, F_0, F_0, F_0, F_0] \ldots \quad (19)$$

For example, for $N = 1024$ channels we need to use $F[10]$ ($2^{10} = 1024$), and so on. In Mathematica, all these operations are achieved and represented by

$$G[N\_] := \text{Partition}[\text{Flatten}[F[\text{Log}[2,N]][[\#;; \ ;; 2]]\&/@\{1,2\}], N] \ . \ N \geq 2 \quad (20)$$

For example, $G[4] = \begin{pmatrix} 1 & 0 & 0 & 0 \\ 1 & 0 & 1 & 0 \\ 1 & 1 & 0 & 0 \\ 1 & 1 & 1 & 1 \end{pmatrix}$, $G[8] = \begin{pmatrix} 1 & 0 & 0 & 0 & 0 & 0 & 0 & 0 \\ 1 & 0 & 1 & 0 & 0 & 0 & 0 & 0 \\ 1 & 0 & 0 & 0 & 1 & 0 & 0 & 0 \\ 1 & 0 & 1 & 0 & 1 & 0 & 1 & 0 \\ 1 & 1 & 0 & 0 & 0 & 0 & 0 & 0 \\ 1 & 1 & 1 & 1 & 0 & 0 & 0 & 0 \\ 1 & 1 & 0 & 0 & 1 & 1 & 0 & 0 \\ 1 & 1 & 1 & 1 & 1 & 1 & 1 & 1 \end{pmatrix}$, ... etc

For an input $u[4]$, the operation $u[4].G[4]$ gives $x[4] = u[4].G[4]$, the codeword $x[4]$ from Figure 19.

In general,

$$x[k] = \text{Mod}[u[k].G[k], 2]. \quad (21)$$

For example,

x2 = Mod[$u[2].F0, 2$] = {Mod[$u_1 + u_2, 2$], Mod[$u_2, 2$]},

$x[4] = \text{Mod}[u[4].G[4], 2] =$

= {Mod[$u_1 + u_2 + u_3 + u_4, 2$], Mod[$u_3 + u_4, 2$], Mod[$u_2 + u_4, 2$], Mod[$u_4, 2$]},

$x[8] = \text{Mod}[u[8].G[8], 2] =$

= {Mod[$u_1 + u_2 + u_3 + u_4 + u_5 + u_6 + u_7 + u_8, 2$], Mod[$u_5 + u_6 + u_7 + u_8, 2$], Mod[$u_2 + u_4 + u_6 + u_8, 2$], Mod[$u_6 + u_8, 2$], Mod[$u_3 + u_4 + u_7 + u_8, 2$], Mod[$u_7 + u_8, 2$], Mod[$u_4 + u_8, 2$], Mod[$u_8, 2$]} .

The codeword $x[N]$ formation is simply created by this procedure for all $N$ of practical use. $x[N]$ supplies the input to the lossy transmission channels $W$. See Figure 1.

This coding is done in the transmission station, before any necessary physical modulation of the signals for transmission through the lossy channels (antenna pathways or network).

It should be emphasized that each $x$ channel contains intertwined contributions from other channels. For example, see $x[8]$ which, evolved during transmission, becomes $y[8]$, a noisy signal.

Continuing the example, on the AWGN channel, $y[8]$ assumes that a Gaussian noise may exchange bits at random places: bit 0 → bit 1 and vice-versa.

## Additive white Gaussian noise channel

### After transmission by a noisy channel: Bit flip due to noise.

Bit flip is a possibility due to signal intensity fluctuations in a channel or possibly due to a phase-flip resulting from the interaction of the carrier signal with obstacles along the signal path.

Upon transmission by the lossy channel, the noiseless codeword vector $\boldsymbol{x}$ is received as the noisy vector $\boldsymbol{y}$ (see Figure 1), where $\boldsymbol{y} = \boldsymbol{x} + \boldsymbol{n}$ and $\boldsymbol{n}$ is the noise added. The number of components in both vectors $\boldsymbol{x}$ and $\boldsymbol{y}$ is the same but it is not a-priori evident which channels lead to a better Mutual Information. Mutual Information can be calculated for each channel if we apply a specific noise model for that transmission channel.

### Mutual Information

Mutual Information can be calculated for each communication cycle or package sent. In this case it reveals the statistical common knowledge between the emitter, represented by $X$ and the receiver, represented by $Y$. By definition,

$$I(X;Y) \equiv H(X) - H(X|Y) \tag{22}$$

$I(X;Y)$ can be rewritten[2]

$$I(X;Y) = \sum_{x,y} P(x,y) \log_2 \frac{P(x,y)}{P(x)P(y)} = I(Y;X) \tag{23}$$

From the Joint probability $P(x, y)$ one can write the Conditional probabilities

$$P(x|y) \equiv \frac{P(x,y)}{P(y)}, \tag{24}$$
$$P(y|x) \equiv \frac{P(x,y)}{P(x)}.$$

Thus

$$I(X;Y) = \sum_{x,y} P(x|y)P(y) \log_2 \frac{P(x|y)P(y)}{P(x)P(y)} = \sum_{x,y} P(x|y)P(y) \log_2 \frac{P(x|y)}{P(x)}, \text{ or}$$
$$I(X;Y) = \sum_{x,y} P(y|x)P(y) \log_2 \frac{P(y,x)}{P(y)}. \tag{25}$$

---

[2] For example, see Eq. (8.27) in *Information Theory, Inference, and Learning Algorithms,* David J.C. MacKay, Cambridge University Press 2003.

$P(y)$ could be given either by $P(y|0)$ or $P(y|1)$, with equal probability of ½. It may be written

$$P(y) = \tfrac{1}{2}(P(y|0) + P(y|1)) = \tfrac{1}{2}, \qquad (26)$$

given that both possibilities are covered by $(P(y|0) + P(y|1))$. Thus

$$I(Y;X) = \sum_{x,y} P(y|x) \log_2(2 \times P(y|x)) \qquad (27)$$

## Probability of error

We may look at Mutual Information for a specific number of channels. The <u>Additive white Gaussian noise channel</u> will be used. One may assume that the Gaussian noise is centered at 0 with variance $\sigma^2$. Bits are mapped as follows: bit 1→ +1 and to bit 0→ −1. The probability for bit +1 or bit -1 is given by

$$p_{\{-1\}^{+1}} = \frac{1}{\sigma\sqrt{2\pi}} e^{-\frac{(y \mp 1)^2}{2\sigma^2}}, \qquad (28)$$

where $y = x + n$. It is essentially dictated by the noise variance $\sigma^2$, a quantity to be experimentally determined. It is clear that $n$ influences $p_{\{-1\}^{+1}}$.

Assuming that the bit signals are transmitted by a carrier with channels modulated by sidebands with close frequency separation, all of them will be affected by the noise in a similar way.

Figure 20 shows $p_{\{-1\}^{+1}}$ for $\sigma = 0.5$.

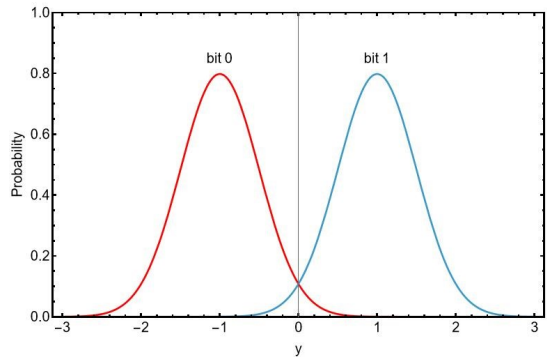

*Figure 20 -* $p_{\{-1\}^{+1}}$ *are shown for the two bits, where the noisy channel has standard deviation $\sigma = 0.5$. The overlap area gives the flipping probability.*

For bit=+1, $y = 1 + n$ and a flip may occur if $y \leq 0$ or $n \leq -1$. Conversely, for bit=-1, $y = -1 + n$ and a flip may occur if $y > 0$ or $n > 1$.

The overlap area for bit=+1 is

$$P_{+1\text{flip}} = \int_{-\infty}^{0} \frac{1}{\sigma\sqrt{2\pi}} e^{-\frac{(y-1)^2}{2\sigma^2}} dy = \frac{1}{2}\text{Erfc}\left[\frac{1}{\sigma\sqrt{2}}\right]. \tag{29}$$

Similarly, the overlap area for bit=-1 is

$$P_{-1\text{flip}} = \int_{0}^{\infty} \frac{1}{\sigma\sqrt{2\pi}} e^{-\frac{(y+1)^2}{2\sigma^2}} dy = \frac{1}{2}\text{Erfc}\left[\frac{1}{\sigma\sqrt{2}}\right]. \tag{30}$$

The total error probability (flip probability) $P_e$ is

$$P_e = P(\text{error}|\text{bit} = +1)P(\text{bit} = +1) + P(\text{error}|\text{bit} = -1)P(\text{bit} = -1) = \tag{31}$$
$$= \frac{1}{2}\big(P(\text{error}|\text{bit} = +1) + P(\text{error}|\text{bit} = -1)\big)$$

or

$$P_e = \frac{1}{2} \times P_{+1\text{flip}} + \frac{1}{2} \times P_{-1\text{flip}} = \text{Erfc}\left[\frac{1}{\sigma\sqrt{2}}\right]. \tag{32}$$

Figure 21 exemplifies the distinct overlaps for different $\sigma$ values.

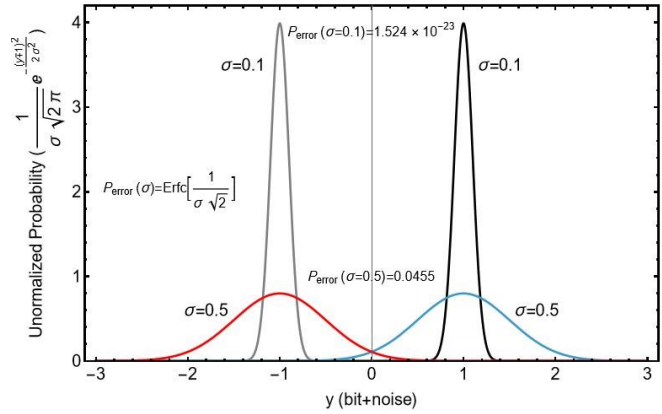

*Figure 21 – Noise effects for cases $\sigma = 0.1$ and $\sigma = 0.5$. The probability for an error (bit flip) is given by $P_{err}(\sigma)$ and directly related to the overlapped area under two curves with the same $\sigma$.*

## Numerical simulations for BEC and AWGN channels

As an example, let us look at the Mutual Information for the received $y[8]$, after transmission of the input $x[8]$. In Polar Coding, channel separation is more evident for many channels, but $x[8] \to y[8]$ uses the same calculation steps for a general case.

$x[8] = \text{Mod}[u[8].G[8], 2] =$

$= \{\text{Mod}[u_1 + u_2 + u_3 + u_4 + u_5 + u_6 + u_7 + u_8, 2], \text{Mod}[u_5 + u_6 + u_7 + u_8, 2],$

$\text{Mod}[u_2 + u_4 + u_6 + u_8, 2], \text{Mod}[u_6 + u_8, 2], \text{Mod}[u_3 + u_4 + u_7 + u_8, 2],$

$\text{Mod}[u_7 + u_8, 2], \text{Mod}[u_4 + u_8, 2], \text{Mod}[u_8, 2]\},$

After transmission, noise will be added and

$$y[8] \leftarrow x[8] + n[8], \tag{33}$$

where $n[8]$ is the noise added by the channel.

### Number of erased channels

In BEC model with $N$ channels and loss $\epsilon$ (for a given distance), the number of erased channels is estimated by

$$\text{Number of erased channels} = \text{Round}[N \times \epsilon]. \tag{34}$$

To find which channels will be erased this simulation uses:

$$\text{Union}\Big[\text{Sort}\Big[\text{RandomInteger}[\{1, n\}, \text{Round}[n \times \epsilon]]\Big]\Big]$$

In the AWGN model the estimated number of affected channels is

$$\text{Number of flipped channels} = \text{Round}\Big[N \times \text{Erfc}\Big[\tfrac{1}{\sigma\sqrt{2}}\Big]\Big]. \tag{35}$$

Figure 22 shows the number of bits affected by noise, $N \times \text{Erfc}\Big[\tfrac{1}{\sigma\sqrt{2}}\Big]$, in the vertical axis, as a function of $N$ and $\sigma$.

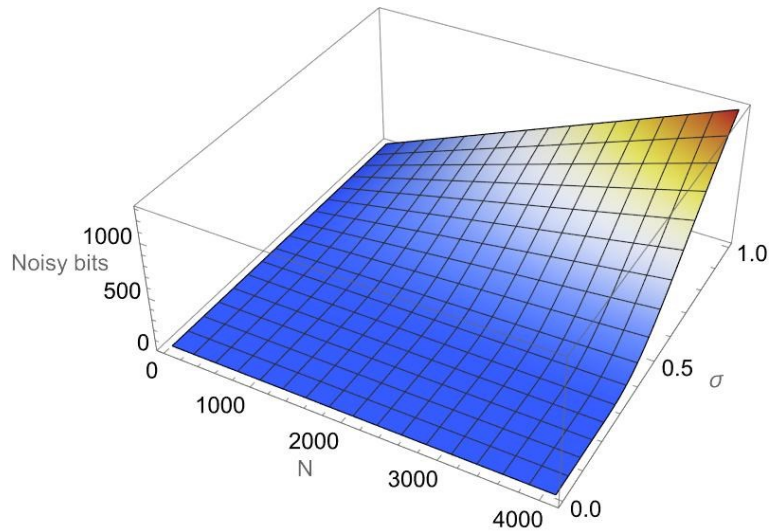

*Figure 22 – Number of bits affected by noise, as a function of the total number of bits N and the noise's standard deviation σ.*

The probability of error in these models, $\epsilon$ and $\text{Erfc}\Big[\tfrac{1}{\sigma\sqrt{2}}\Big]$, contains the idea of fixed conditions within the time scale taken by signal transmission. For turbulent channels, with both short and long-time condition variations, more elaborate calculations can be produced [4].

## Linear Decoding

Decoding operations happen at the RX station. A program was written in Mathematica to perform a linear decoding (LD). Linear decoding is **not** successive-cancellation decoding as described by Arikan. Explanatory notes are inserted at each program step described below. The program efficiency for a full decoding in a **<u>first</u>** trial is exemplified for a range of $\epsilon$ of $\sigma$ values. It is shown that linear decoding in a first trial is efficient at low noisy channels. The percentage of "fail" in a first trial, as the channel noise increases, will be presented.

### "Do" cycle program

Section "Encoding protocol" describes Arikan's polar coding protocol. This encoding protocol is taken as the first, or fixed part, of the decoding program, before a Do cycle is applied for LD. Repetition of decoding allows a statistical analysis of the linear-decoding process.

## Encoding protocol (Arikan's)

To enable the reader's understanding of format information and definitions, Arikan's main coding steps are listed here:

| |
|---|
| $F_0 = \begin{pmatrix} 1 & 0 \\ 1 & 1 \end{pmatrix}$ ;(* seed XoR *) |
| $F^{\otimes n} \coloneqq \text{Nest}[\text{KroneckerProduct}[\#,\#]\&, F_0, n-1]$; (*nested $F_0$*) |
| $G[N\_] = K[N] = F^{\otimes N}$; |
| $s[N\_] \coloneqq \text{Table}[s_k, \{k,1,N\}]; R[N\_] \coloneqq \text{Flatten}[s[N][[\#;;\,;;2]]\&/@\{1,2\}]$; <br> (*$R[N]$ is the permutation operator*) |
| $u[N\_] \coloneqq \text{Table}[u_k, \{k,1,N\}]; uv[N\_] \coloneqq \text{Table}[u[k], \{k,1,N\}]$; <br> (*message bits to be inserted at the beginning*) |
| $Ru[N\_] \coloneqq \text{Flatten}[uv[N][[\#;;\,;;2]]\&/@\{1,2\}]$  (* for the message bits *) |
| $x[N\_] \coloneqq \text{Flatten}[Ru[N],1].K[Log[2,N]]$; <br> (*this is the coded message, before the noisy channel. $K[Log[2,N]]$ transforms the bit input onto the polar coding=coded word before transmission*) |
| $uvr[N\_] \coloneqq \text{Table}[ur[k], \{k,1,N\}]$; (* message bits in general form to be recovered at receiver *) |
| $Ruv[N\_] \coloneqq \text{Flatten}[uvr[N][[\#;;\,;;2]]\&/@\{1,2\}]$ |
| $y[N\_] \coloneqq \text{Flatten}[Ruv[N],1].K[Log[2,N]]$; <br> (* this is $y[N]$ at the receiver, but with noise still to be added, in each cycle of Do*) |

## Linear Decoding protocol (Gaussian flip)

The linear decoding protocol is written for repeated decoding, in a Do cycle, with distinct noise at each cycle. This generates the probability of a successful decoding in a first trial or its failure.

```
Do[tcc=ToCharacterCode["Write A Write BC"];
 id=IntegerDigits[tcc,2,8];Clear[e];Clear[eb];
 σ=0.005;n=256;{t0=TimeUsed[];
 mssg=Flatten[id] ;
 mL=Length[mssg];
 fb=Table[f,{k,2,n,2}](*frozen elements on even indexes*);
 ujoined=Join[fb,mssg] ;
 FullMessage=Table[u[k]=ujoined[[k]],{k,1,Length[ujoined]}](*defining u[k] elements of the   full message*);
```

```
ri=Union[Sort[RandomInteger[{1,n},Round[n*Erfc[ 1/(σ√2 ) ]]]]](*noise elements*);
tri=Table[ri[[k]]->e,{k,1,Length[ri]}](* transformation table for noise elements*);
ttri=Table[tri[[k,1]],{k,1,Length[tri]}]; (*indexes where flip occurs *);
tf=Table[{ttri[[k]],BitFlip[x[256][[ttri[[k]]]],0]},{k,1,Length[ttri]}];
tflip=Flatten[Table[{tf[[k,1]]->tf[[k,2]]},{k,1,Length[tf]}]] (* flip rule to apply ahead *);
yef[256]=ReplacePart[x[256],tflip](*received sequence yef with flip positions*);
uvr[N_]:=Table[ur[k],{k,1,N}](*ur symbols to be decoded at the end*);
ye[256]:=ReplacePart[y[256],tflip];
KnownElements=Table[ur[k]=f,{k,1,n/2}](*f is a known element by sender and receiver*);
ye[256] (*ye modified by f*);
tRedui=TimeUsed[];
Redu=Reduce[yef[256]==ye[256],uvr[256][[129;;256]]]; (*Solving for ur[k] of interest*)
tReduf=TimeUsed[];
f==0 (*to impose f value on Redu if needed :*);
tRedu2=Table[Redu[[k,2]],{k,1,Length[Redu]}];
(*tRedu3=Table[{k+64,ursol[[k+64]]=Redu[[k,2]]},{k,1,Length[Redu]}];*);
tRedu3=Table[{k+128,Redu[[k,2]]},{k,1,Length[Redu]}];
 TestMssg=TrueQ[mssg==tRedu2](*testing
solution*);pstRedu=Partition[Table[tRedu2[[k]],{k,1,Length[tRedu2]}],8](*Writing decoded bytes*);
 nDecoded=Table[{k,FromDigits[pstRedu[[k]],2]},{k,1,Length[pstRedu]}](* FORM with k, message in
character codes *);
DecodedMessage=DecodedMessage=Table[FromCharacterCode[nDecoded[[k,2]]],{k,1,Length[nDecoded]}]
(* plaintext in list format *);
 nDecoded2=Table[FromDigits[pstRedu[[k]],2],{k,1,Length[pstRedu]}](* FORM without k, message in
character codes *);
 DecodedMessage2=FromCharacterCode[nDecoded2] ;
tF=TimeUsed[];{yef[256],ye[256]}>>>" FilePath \\N128Sig0p05smax100_Y.dat";
 {n,σ,s,tRedu3,"","",tF-t0,tReduf-
tRedui,TestMssg,tf,DecodedMessage,DecodedMessage2,Date[]}>>>"FilePath\\N128Sig0p05smax100.dat"
 },{s,1,100}]
t2=TimeUsed[];t2-t1
```

## Linear Decoding protocol (Binary Erasure)

Repeated decoding in distinct cycles and different noise within each cycle.

```
Do[tcc=ToCharacterCode["Write A Write BC"];
 id=IntegerDigits[tcc,2,8];Clear[e];Clear[eb];
 ϵ=0.1;n=256;{t0=TimeUsed[];
 mssg=Flatten[id] ;
 mL=Length[mssg];
 fb=Table[f,{k,2,n,2}](*frozen elements on even indexes, or Table[f,{k,1,n/2}] for odd and even indexes*);
 ujoined=Join[fb,mssg] ;
 FullMessage=Table[u[k]=ujoined[[k]],{k,1,Length[ujoined]}](*defining u[k] elements of the  full message*);
 ri=Union[Sort[RandomInteger[{1,n},Round[n* ϵ]]]](*noise elments*);
  tri=Table[ri[[k]]->eras,{k,1,Length[ri]}](* transformation table for noise elements*);
```

```
  ttri=Table[tri[[k,1]],{k,1,Length[tri]}]; (*indexes where flip occurs *);
  tf=Table[{ttri[[k]],eras},{k,1,Length[ttri]}];
  teras=Flatten[Table[{tf[[k,1]]->tf[[k,2]]},{k,1,Length[tf]}]] (* erase rule to apply ahead *);
  yef[256]=ReplacePart[x[256],teras](*received sequence yef with erased positions*);
  uvr[N_]:=Table[ur[k],{k,1,N}](*ur symbols to be decoded at the end*);
  ye[256]:=ReplacePart[y[256],teras];
  KnownElements=Table[ur[k]=f,{k,1,n/2}](*f is a known element by sender and receiver*);
  ye[256] (*ye modified by f*);
  tRedui=TimeUsed[];
  Redu=Reduce[yef[256]==ye[256],uvr[256][[129;;256]]]; (*Solving for ur[k] of interest*)
  tReduf=TimeUsed[];
  f==0 (*to impose f value on Redu if needed :*);
  tRedu2=Table[Redu[[k,2]],{k,1,Length[Redu]}];
  (*tRedu3=Table[{k+64,ursol[[k+64]]=Redu[[k,2]]},{k,1,Length[Redu]}];*);
  tRedu3=Table[{k+128,Redu[[k,2]]},{k,1,Length[Redu]}];
  eras=Nothing;
  TestMssg=TrueQ[mssg==tRedu2](*testing
solution*);pstRedu=Partition[Table[tRedu2[[k]],{k,1,Length[tRedu2]}],8](*Writing decoded bytes*);
  nDecoded=Table[{k,FromDigits[pstRedu[[k]],2]},{k,1,Length[pstRedu]}](* FORM with k, message in
character codes *);
DecodedMessage=DecodedMessage=Table[FromCharacterCode[nDecoded[[k,2]]],{k,1,Length[nDecoded]}]
(* plaintext in list format *);f=0;
  DecodedMessage;
  nDecoded2=Table[FromDigits[pstRedu[[k]],2],{k,1,Length[pstRedu]}](* FORM without k, message in
character codes *);
  DecodedMessage2=FromCharacterCode[nDecoded2] ;
tF=TimeUsed[];{yef[256],ye[256]}>>>" File Path \\N256Eps0p1smax100BEC_Y1.dat";
  {n,ϵ,s,tRedu3,tF-t0,tReduf-
tRedui,TestMssg,DecodedMessage,DecodedMessage2,wrongOVERtotal,Date[]}>>>"File
Path\\N256Eps0p1smax100BEC1.dat"
  },{s,1,100}]
t2=TimeUsed[];t2-t1
```

## Linear Decoding examples – BEC frozen on odd and even indexes

Table 1 shows the Average Decoding time and the number of failing trials. For each line of Table 1, 100 repetitions were performed.

*Table 1 – Average Decoding time (=Reduce time) and number of failing trials divided by the total number of transmission repetitions.*

| Number of bits | ϵ | # Decoding | Run time (s) | Reduce time (s) | Decoded Message / List | Decoded Message / String | FailsOVERtotal | DateTime |
|---|---|---|---|---|---|---|---|---|
| 256 | 0.001 | 1 | 0.078 | 0.062 | {W, r, i, t, e, , A, , W, r, i, t, e, , B, C} | Write A Write BC | 0 | {2025, 7, 7, 15, 17, 19.3470185} |
| 256 | 0.005 | 1 | 0.063 | 0.047 | {W, r, i, t, e, , A, , W, r, i, t, e, , B, C} | Write A Write BC | 0 | {2025, 7, 7, 15, 31, 51.2816051} |
| 256 | 0.01 | 1 | 0.11 | 0.11 | {W, r, i, t, e, , A, , W, r, i, t, e, , B, C} | Write A Write BC | $\frac{1}{25}$ | {2025, 7, 7, 15, 35, 31.4404037} |
| 256 | 0.02 | 1 | 0.062 | 0.047 | {W, r, i, t, e, , A, , W, r, i, t, e, , B, C} | Write A Write BC | $\frac{3}{50}$ | {2025, 7, 7, 15, 37, 50.3908344} |
| 256 | 0.03 | 1 | 0.079 | 0.063 | {W, r, i, t, e, , A, , W, r, i, t, e, , B, C} | Write A Write BC | $\frac{1}{10}$ | {2025, 7, 7, 15, 39, 50.5105753} |
| 256 | 0.04 | 1 | 0.094 | 0.078 | {W, r, i, t, e, , A, , W, r, i, t, e, , B, C} | Write A Write BC | $\frac{6}{25}$ | {2025, 7, 7, 15, 41, 51.3802600} |
| 256 | 0.05 | 1 | 0.078 | 0.047 | {W, r, i, t, e, , A, , W, r, i, t, e, , B, C} | Write A Write BC | $\frac{7}{25}$ | {2025, 7, 8, 13, 54, 29.2193273} |
| 256 | 0.06 | 1 | 0.062 | 0.047 | {W, r, i, t, e, , A, , W, r, i, t, e, , B, C} | Write A Write BC | $\frac{31}{100}$ | {2025, 7, 8, 13, 57, 10.4844650} |
| 256 | 0.07 | 1 | 0.11 | 0.094 | {W, r, i, t, e, , A, , W, r, i, t, e, , B, C} | Write A Write BC | $\frac{51}{100}$ | {2025, 7, 8, 14, 2, 38.9472831} |
| 256 | 0.08 | 1 | 0.125 | 0.11 | {W, r, i, t, e, , A, , W, r, i, t, e, , B, C} | Write A Write BC | $\frac{14}{25}$ | {2025, 7, 8, 14, 4, 57.0816718} |
| 256 | 0.09 | 2 | 0.093 | 0.062 | {W, r, i, t, e, , A, , W, r, i, t, e, , B, C} | Write A Write BC | $\frac{29}{50}$ | {2025, 7, 8, 14, 7, 1.6842282} |
| 256 | 0.1 | 1 | 0.079 | 0.063 | {W, r, i, t, e, , A, , W, r, i, t, e, , B, C} | Write A Write BC | $\frac{4}{5}$ | {2025, 7, 8, 14, 9, 24.1681425} |
| 256 | 0.15 | 2 | 0.094 | 0.062 | {W, r, i, t, e, , A, , W, r, i, t, e, , B, C} | Write A Write BC | $\frac{23}{25}$ | {2025, 7, 8, 14, 11, 8.7396616} |

Table 1 exemplifies the number of decoding failures as a function of the BEC loss $\epsilon$, and shows the decoding time for each case. The average Reduce time per bit = 0.27ms.

## Linear Decoding examples – Gaussian flip frozen on odd and even indexes

For each line of Table 1, 100 repetitions were performed.

*Table 2 - Decoding time (=Reduce time) and number of failing trials divided by the total number of transmission repetitions.*

| Number of bits | σ | ♯ Decoding | Run time (s) | Reduce time (s) | Decoded Message / List | Decoded Message / String | FailsOVERtotal | DateTime |
|---|---|---|---|---|---|---|---|---|
| 1024 | 0.01 | 1 | 0.39 | 0.343 | {D, õ, 6, ß, , M, (, , , », ?, 5, Ô, þ, , ∅, \, Ü, à, Ê, ò, ½, ¼, ", ¹, <, *, i, , ·, ∅, Ä, , e, x, N, ä, , ·, È, 3, ∅, ú, , , , , , ý, !, ê, Ó, , d, ý, ×, ∅, ´, Þ, 4, , î, p, y, } | Dõ6ß M(,»?5Ôþ∅\ÜàÊò½¼"¹<*i. ∅ÄexNä·È3∅ú ý!êÓdý×∅´Þ4îpy | 0 | {2025, 7, 7, 12, 2, 48.7372378} |
| 1024 | 0.05 | 1 | 0.578 | 0.359 | {D, õ, 6, ß, , M, (, , , », ?, 5, Ô, þ, , ∅, \, Ü, à, Ê, ò, ½, ¼, ", ¹, <, *, i, , ·, ∅, Ä, , e, x, N, ä, , ·, È, 3, ∅, ú, , , , , , ý, !, ê, Ó, , d, ý, ×, ∅, ´, Þ, 4, , î, p, y, } | Dõ6ß M(,»?5Ôþ∅\ÜàÊò½¼"¹<*i. ∅ÄexNä·È3∅ú ý!êÓdý×∅´Þ4îpy | 0 | {2025, 7, 7, 11, 46, 5.3509741} |
| 1024 | 0.1 | 1 | 0.547 | 0.312 | {D, õ, 6, ß, , M, (, , , », ?, 5, Ô, þ, , ∅, \, Ü, à, Ê, ò, ½, ¼, ", ¹, <, *, i, , ·, ∅, Ä, , e, x, N, ä, , ·, È, 3, ∅, ú, , , , , , ý, !, ê, Ó, , d, ý, ×, ∅, ´, Þ, 4, , î, p, y, } | Dõ6ß M(,»?5Ôþ∅\ÜàÊò½¼"¹<*i. ∅ÄexNä·È3∅ú ý!êÓdý×∅´Þ4îpy | 0 | {2025, 7, 7, 14, 47, 49.5976201} |
| 1024 | 0.2 | 1 | 0.563 | 0.391 | {D, õ, 6, ß, , M, (, , , », ?, 5, Ô, þ, , ∅, \, Ü, à, Ê, ò, ½, ¼, ", ¹, <, *, i, , ·, ∅, Ä, , e, x, N, ä, , ·, È, 3, ∅, ú, , , , , , ý, !, ê, Ó, , d, ý, ×, ∅, ´, Þ, 4, , î, p, y, } | Dõ6ß M(,»?5Ôþ∅\ÜàÊò½¼"¹<*i. ∅ÄexNä·È3∅ú ý!êÓdý×∅´Þ4îpy | 0 | {2025, 7, 7, 12, 10, 19.9885557} |
| 1024 | 0.3 | 1 | 0.672 | 0.36 | {D, õ, 6, ß, , M, (, , , », ?, 5, Ô, þ, , ∅, \, Ü, à, Ê, ò, ½, ¼, ", ¹, <, *, i, , ·, ∅, Ä, , e, x, N, ä, , ·, È, 3, ∅, ú, , , , , , ý, !, ê, Ó, , d, ý, ×, ∅, ´, Þ, 4, , î, p, y, } | Dõ6ß M(,»?5Ôþ∅\ÜàÊò½¼"¹<*i. ∅ÄexNä·È3∅ú ý!êÓdý×∅´Þ4îpy | 0 | {2025, 7, 7, 12, 50, 10.2648953} |
| 1024 | 0.4 | 1 | 1.687 | 0.297 | {D, õ, 6, ß, , M, (, , , », ?, 5, Ô, þ, , ∅, \, Ü, à, Ê, ò, ½, ¼, ", ¹, <, *, i, , ·, ∅, Ä, , e, x, N, ä, , ·, È, 3, ∅, ú, , , , , , ý, !, ê, Ó, , d, ý, ×, ∅, ´, Þ, 4, , î, p, y, } | Dõ6ß M(,»?5Ôþ∅\ÜàÊò½¼"¹<*i. ∅ÄexNä·È3∅ú ý!êÓdý×∅´Þ4îpy | $\frac{1}{25}$ | {2025, 7, 7, 12, 58, 5.2111590} |
| 1024 | 0.5 | 5 | 6.734 | 0.422 | {D, õ, 6, ß, , M, (, , , », ?, 5, Ô, þ, , ∅, \, Ü, à, Ê, ò, ½, ¼, ", ¹, <, *, i, , ·, ∅, Ä, , e, x, N, ä, , ·, È, 3, ∅, ú, , , , , , ý, !, ê, Ó, , d, ý, ×, ∅, ´, Þ, 4, , î, p, y, } | Dõ6ß M(,»?5Ôþ∅\ÜàÊò½¼"¹<*i. ∅ÄexNä·È3∅ú ý!êÓdý×∅´Þ4îpy | $\frac{63}{100}$ | {2025, 7, 7, 13, 10, 28.0267446} |
| 1024 | 0.6 | 27 | 13.047 | 0.594 | {D, õ, 6, ß, , M, (, , , », ?, 5, Ô, þ, , ∅, \, Ü, à, Ê, ò, ½, ¼, ", ¹, <, *, i, , ·, ∅, Ä, , e, x, N, ä, , ·, È, 3, ∅, ú, , , , , , ý, !, ê, Ó, , d, ý, ×, ∅, ´, Þ, 4, , î, p, y, } | Dõ6ß M(,»?5Ôþ∅\ÜàÊò½¼"¹<*i. ∅ÄexNä·È3∅ú ý!êÓdý×∅´Þ4îpy | $\frac{99}{100}$ | {2025, 7, 7, 13, 53, 30.4501132} |

Table 2 exemplifies the number of decoding failures as a function of Gaussian standard deviation $\sigma$, and shows the decoding time for each case. The average Gaussian Flip Reduce time per bit = 0.38ms.

For the Gaussian-flip channel Figure 23 shows the number of failing decoding in a single trial.

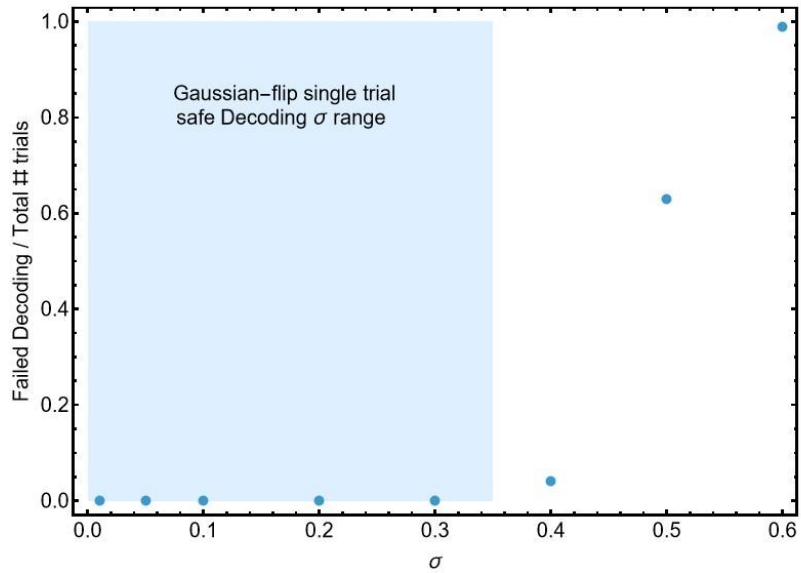

*Figure 23 - Failed decoding due to noise in the Gaussian flip mode as a function of the Standard Deviation σ. The blue shaded area is the region where decoding is achieved with no errors in a single decoding trial.*

## "Failed" decoding plots for Gaussian and BEC channels

For the BEC channel Figure 24 shows the number of failing decoding in a single trial.

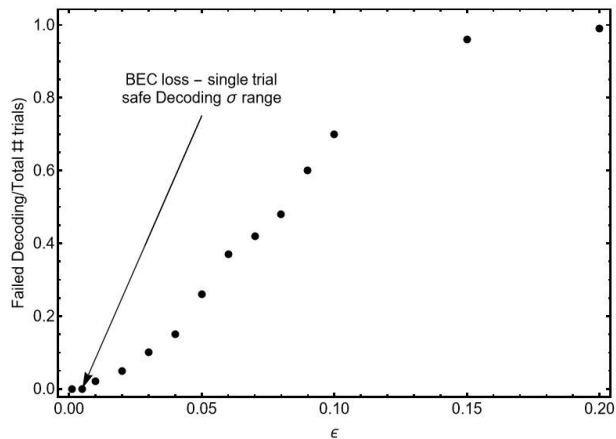

*Figure 24 - Failed decoding due to noise in the BEC model, as a function of the Probability of error ϵ. The range of ϵ for decoding in a first trial is small.*

The connection between σ and the probability of error for the Gaussian model is shown in Figure 25

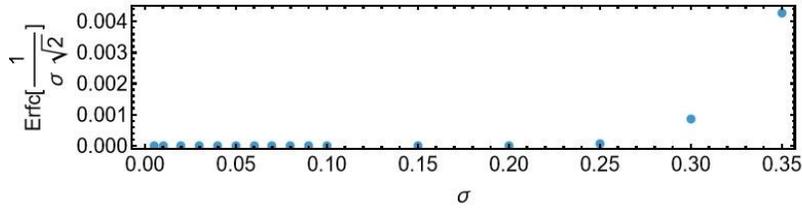

*Figure 25 - Connection between σ and the probability of error for the Gaussian model.*

It is possible to numerically find the values of $\sigma$ satisfying the condition $\text{Erfc}\left[\frac{1}{\sigma\sqrt{2}}\right] = \epsilon$. See Figure 26. For example, for $\epsilon \cong 0.5$, a value $\sigma = 1.48$ is obtained.

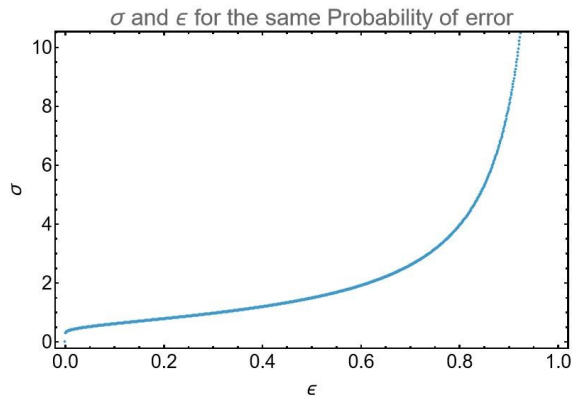

*Figure 26 - Values of σ giving the same Probability of error as ϵ.*

Increasing the number of bits (or codewords) lead to similar results.

## Repeated transmission

Figure 23 and Figure 24 shows linear decoding failure cases as a function of $\sigma$ or $\epsilon$. The simple existence of decoding failure and success for the same data content indicated that the noise action on a random position $y_i$ decides the importance of that particular $y_i$ in solving the linear equations for decoding. This also implies that a repetition of that transmission changes the noise impact on $y_i$: noise may affect another channel $y_k$ ($k \neq i$) with less importance for that instance of decoding. This suggests that a single repetition of a linear decoding procedure can be applied for some ranges of $\sigma$ to achieve a successful decoding.

The number of transmissions for successful decoding of a data set can be calculated from information taken from Figure 23 and Figure 24, as # of transmissions $\sim \frac{1}{1-\text{Failed Decoding}}$. Figure 27 and Figure 28 show the results.

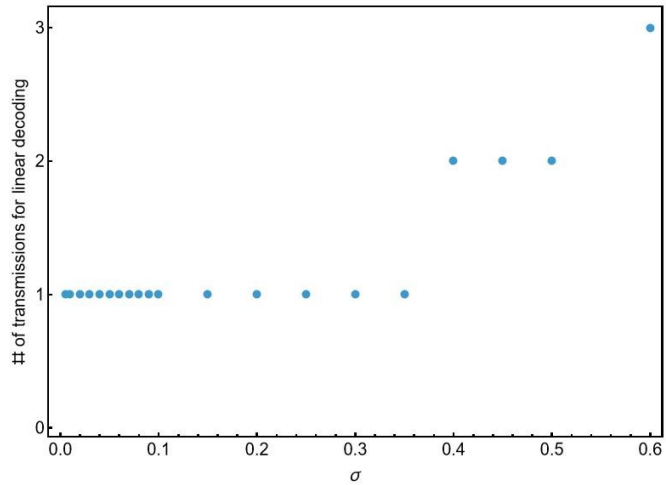

*Figure 27 – Number of transmissions (repetitions) to achieve a successful decoding as function of σ.*

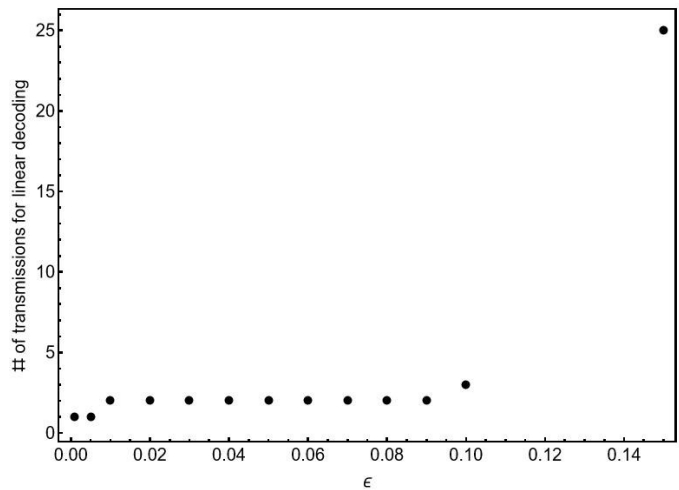

*Figure 28 – Number of transmissions to achieve a successful decoding as function of $\epsilon$.*

Nonlinear decoding can be applied for low noise channels, such as optical channels with controlled absorption. Estimates are given for ranges where transmission repetitions may be needed to guarantee successful decoding. This procedure differs from Arikan's probabilistic successive decoding. Practical comparisons between the two processes have yet to be made.

Successful decoding for random data can be verified in repeated emissions using a known data sequence together with the unknown message. Successful decoding of the known data sequence may result in an acceptable probabilistic guarantee about the correct decoding of the message.

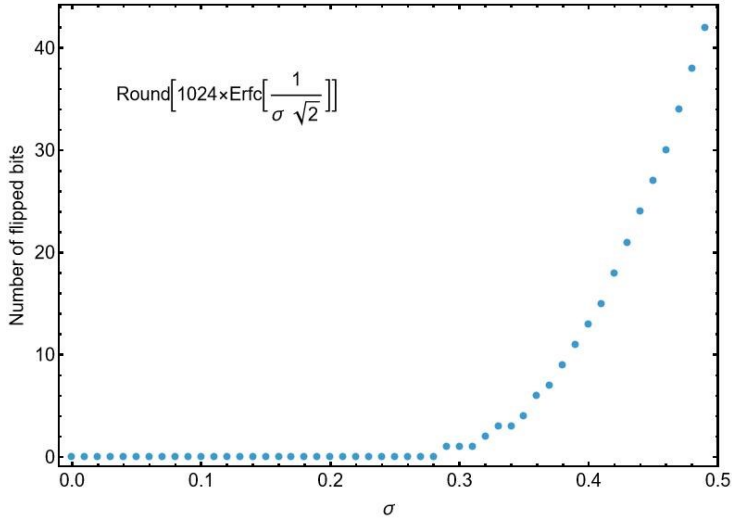

*Figure 29 – Average number of randomly flipped bits as a function of σ. For low values up to σ~0.35 a single transmission repetition decodes the information (see Figure 27), because the number of flipped bits is low. With a few repetitions a successful decoding is achieved for 10-30 flipped bits.*

### Decoding time

Arikan's and similar decoding schemes have a complexity of successive cancellation with a $N$ dependence $O(N \log N)$, less demanding than $10^{-6}N^2$ as shown in Figure 15. Another important parameter for decoding is a time-per-bit in the decoding phase.

Arikan reports in "Polar Coding-Status and Prospects", *The IEEE International Symposium on Information Theory (ISIT'2011)*, slide 122, that Polar Decoding for $N \sim 770$, takes approximately 1ms per codeword.

In the case of $N = 1024$, the linear decoding presented in this note was $(0.60 \pm 0.09)$ms for single trial decoding. Even with a repetition, as indicated in Figure 27, the decoding time was similar in magnitude to Arikan's successive decoding.

The linear decoding time was obtained at the step of solving the set of linear equations

**Redu=Reduce[yef[256]==ye[256],uvr[256][[129;;256]]],**  (36)

as written in the "Linear Decoding protocol (Gaussian flip)".

### Conclusions

Probabilistic Successive Decoding protocols have been developed in software and hardware and adopted for 5G telecommunication around the world. This note explores the alternative of decoding by directly solving the set of linear equations connecting the codewords $x$ and the receiver signals $y$ corrupted by noise in the transmission stage. For this task, Mathematica language programming was used. Decoding by solving linear equations has a complexity $O(N^2)$, and can be explored by using FPGA and ASIC technologies for increased speeds, and comparisons to Successive Decoding. Polar coding techniques may also apply to deterministic quantum-resistant protocols (see https://csrc.nist.gov/projects/post-quantum-cryptography) and to <u>non-deterministic</u> quantum-resistant protocols [17].